Comment on "Visualization of the interplay between high-temperature superconductivity, the pseudogap and impurity resonances" (K. Chatterjee *et al.*, *Nature Phys.* **4**, 108 (2008))

In a recent letter, Chatterjee *et al.* reported tunnelling data obtained below and above the critical temperature ($T_c$ = 15 K) by STM in overdoped $Bi_{2-y}Pb_ySr_2CuO_{6+x}$ (Bi2201) and some interpretations of the data[1]. The main experimental findings of their work are observations of native-impurity resonances which coexist spatially with the superconducting (SC) gap below $T_c$ and survive unchanged on warming through $T_c$. No doubts, the data in the letter are an important piece of information for assembling the high-$T_c$ SC jigsaw puzzle; however, there are inaccuracies in the discussion. In addition, I propose another interpretation of the data.

They wrote "Experimentally, a number of recent results suggest that the pseudogap not only exist above $T_c$ but also coexist with the SC gap below $T_c$ (refs 17-20)." The statement is inaccurate. The fact of the coexistence of the SC gap and the pseudogap below $T_c$ in cuprates was found a long time ago. In heat-capacity[2] and NMR[3] measurements, this was deduced for YBCO 14 years ago. To distinguish, let us call it as the NMR pseudogap. In tunnelling measurements, the coexistence of the SC gap and the *tunnelling* pseudogap below $T_c$ in Bi2212 was pointed out, at least, 7 years ago[4,5]. Further in the letter, the authors use some findings for the NMR pseudogap in order to explain their data. This can be erroneous. Not considering a contribution of incoherent Cooper pairs above $T_c$ in cuprates, if such exists, a direct comparison of the two normal-state pseudogaps obtained by NMR *and* by tunnelling is not straightforward. On the phase diagram of cuprates, the doping dependences of the two pseudogaps differ[3,6,7]. This fact can be easily understood in the framework of a low-$T$ phase-separation scenario. In addition, it is worth noting that the two techniques—NMR and tunnelling—probe different parts of the sample—bulk and surface, correspondently. And, in phase-separated materials, the two techniques can be phase-selective, i.e. they may probe different phases (e.g. tunnelling – a conducting phase, and NMR – an insulating one).

In addition to the explanations of the tunnelling data[1], there is, at least, one more. Earlier, analysis of tunnelling data allowed one to conclude about the origin of quasiparticles in cuprates: it was found that quasiparticles in Bi2212 are soliton-like excitations[5-7]. Generally speaking, solitonic states appear inside of a normal-state gap, and their in-gap position is determined by the system[8,6]. In a system, the density of solitons cannot be large (there is a maximum value) and, and if the solitons are dynamic (and not part of the SC condensate), it is virtually impossible to observe them by a "slow" experimental technique. However, pinned by a defect or by an impurity, the soliton-like excitations can be easily detected near the defect/impurity by a "slow" technique. That is exactly what might occur in the Bi2201[1]. Tunnelling is a "slow" technique: one spectrum is recorded, at best, during tens of microseconds (electron-spin relaxation in solids is of the order of a few nanoseconds). And the last, the authors of the letter did not notice that their data may have a lot in common with resonances observed on YBCO chains[9].

A. Mourachkine
Cavendish Laboratory, University of Cambridge, J.J. Thomson Ave., Cambridge CB3 0HE, UK